# Bipolar Conduction is the Origin of the Electronic Transition in Pentatellurides: Metallic vs. Semiconducting Behavior

P. Shahi[1], D. J. Singh[2*], J. P. Sun[1], L. X. Zhao[1], G. F. Chen[1], J.-Q. Yan[3], D. G. Mandrus[3,4], and J.-G. Cheng[1*]

[1.] Beijing National Laboratory for Condensed Matter Physics and Institute of Physics, Chinese Academy of Sciences, Beijing 100190, China
[2.] Department of Physics and Astronomy, University of Missouri, Columbia, MO 65211-7010 USA
[3.] Materials Science and Technology Division, Oak Ridge National Laboratory, Oak Ridge, TN 37831, USA
[4.] Department of Materials Science and Engineering, University of Tennessee, Knoxville, TN 37996, USA

[*]E-mails: singhdj@missouri.edu and jgcheng@iphy.ac.cn

## Abstract

The pentatellurides, $ZrTe_5$ and $HfTe_5$ are layered compounds with one dimensional transition-metal chains that show a never understood temperature dependent transition in transport properties as well as recently discovered properties suggesting topological semimetallic behavior. Here we show that these materials are semiconductors and that the electronic transition is due to a combination of bipolar effects and different anisotropies for electrons and holes. We report magneto-transport properties for two kinds of $ZrTe_5$ single crystals grown with the chemical vapor transport (S1) and the flux method (S2), respectively. These have distinct transport properties at zero field: the S1 displays a metallic behavior with a pronounced resistance peak and a sudden sign reversal in thermopower at approximately 130 K, consistent with previous observations of the electronic transition; in strikingly contrast, the S2 exhibits a semiconducting-like behavior at low temperatures and a positive thermopower over the whole temperature range. For both samples, strong effects on the transport properties are observed when the magnetic field is applied along the orthorhombic *b* and *c* axes. Refinements on the single-crystal X-ray diffraction and the energy dispersive spectroscopy analysis revealed the presence of noticeable Te-vacancies in the sample S1, confirming that the widely observed anomalous transport behaviors in pentatellurides actually take place in the Te-deficient samples. Electronic structure calculations show narrow gap semiconducting behavior, with different transport anisotropies for holes and electrons. For the degenerately doped *n*-type samples, our transport calculations can result in a resistivity peak and crossover in thermopower from negative to positive at temperatures close to that observed experimentally. Our present work resolves the longstanding puzzle regarding the anomalous transport behaviors of pentatellurides, and also resolves the electronic structure in favor of a semiconducting state.

# Introduction

The pentatellurides, ZrTe$_5$ and HfTe$_5$, are layered materials,[1, 2] with a high concentration of the heavy *p*-element Te, and which show a remarkable electronic transition as a function of temperature.[3] These compounds occur in an orthorhombic space group *Cmcm* with layers stacked along the *b*-axis direction, and the chains of Zr or Hf atoms running along the *a*-axis (as depicted in Fig. S1). The crystals generally grow with a needle- or ribbon-like morphology. Transport measurements are normally made along the *a*-axis direction. Importantly, both compounds are reported to show a strong peak in resistivity, $\rho(T)$, accompanied by a change in sign of the thermopower, $S(T)$, from electron-like to hole-like as $T$ is increased. This takes place at ~130–150 K for ZrTe$_5$ and ~50-80 K for HfTe$_5$ samples grown by similar methods.[3] It was observed early on that this transition does not show a sharp structure in resistivity, but rather has a continuous behavior[3] and that there is no evident structural distortion around the transition as in for example a charge density wave.[4] It was also noted that these compounds have rather high values of thermopower at the peak, making them of interest as potential thermoelectric materials.[5] Finally, it is noteworthy that samples grown by different methods show qualitatively similar behavior but that the temperature at which the peak occurs may differ. For example, recent flux grown ZrTe$_5$ samples show a peak at ~60 K.[6] In any case, possible causes for this behavior were advanced. These include density waves,[3] inconsistent with the diffraction and high magnetic field data,[4] polaronic models,[7] which are, however, apparently inconsistent with the good low temperature conduction, or a semimetal-semiconductor phase transition,[8] or a temperature-induced Lifshitz transition[9, 10]. More recently, these compounds have become of interest as topological materials whose low energy electronic structure is controlled by spin orbit.[11] Nevertheless, it is currently under hot debate whether they are topological insulators or Dirac semimetals.[6, 12-18]

In order to resolve these issues, we have performed a comprehensive study on the transport properties for two kinds of ZrTe$_5$ crystals grown with the chemical vapor transport and the flux method, denoted as S1 and S2 hereafter, respectively. These are the two main techniques that have been used to grow pentatelluride single crystals. Our results on the S1 sample agree well with the previous observations, *i.e.* the metallic behavior with a resistance peak and a sign reversal of thermopower, but the S2 sample displays semiconducting-like behavior with positive thermopower in the whole temperature range. These distinct behaviors are found to originate from the quite different Zr: Te ratios, *i.e.* 1: 4.60 ± 0.20 for S1 versus 1: 4.98 ± 0.17 for S2, based on the energy dispersive spectroscopy measurements. The presence of Te vacancies in S1 was also verified by the refinement on the single-crystal X-ray diffraction data. Te vacancies are an expected *n*-type dopant. Electronic structure calculations for stoichiometric material show narrow gap semiconducting behavior, with a two dimensional electronic structure having opposite high conductivity directions for electrons and holes. Transport calculations show that with *n*-type doping a resistivity peak at temperatures consistent with experiment is obtained. This peak is accompanied by a change in sign from *n*-type to *p*-type thermopower, and has its origin in the onset of bipolar conduction. Simple semiconducting behavior is obtained without doping. The effect of magnetic field can be explained by a standard field dependent reduction of conductivity perpendicular to the field direction. Thus, our present work not only resolves the long-standing puzzle regarding the anomalous transport behaviors, and also resolves question of the electronic structure in favor of a semiconducting state.

## Experimental Details

$ZrTe_5$ single crystals used in the present study were grown by the chemical vapor transport (CVT) for the S1 samples[16] and the Te-flux method for the S2 samples[18]. We employed iodine ($I_2$) as the transport agent during the CVT crystal growth. Zr (powder 99.2%, Hf nominal 4.5%) and Te (powder, 5N) were mixed in the molar ratio of Zr: Te = 1: 5.5 and then sealed in a quartz ampoule with $I_2$ (7 mg/mL). The ampoule was placed in a two-zone furnace. Typical temperature gradient from 480 °C to 400 °C was applied. After two month, long ribbon-shaped single crystals were obtained. On the other hand, we used a Canfield crucible set (CCS)[19] to grow $ZrTe_5$ crystals out of Te flux.[18] Zr slug (99.95%, Hf nominal 3%) and Te shots (5N) in an atomic ratio of 1: 49 were loaded into the CCS and then sealed in a silica ampoule under vacuum. The sealed ampoule was heated to 1000 °C and kept for 12 h to homogenize the melt, furnace cooled to 650°C, and then cooled down to 460 °C in 60 h. $ZrTe_5$ crystals were isolated from Te flux by centrifuging at 460 °C. Typical $ZrTe_5$ (S2) crystals are about 10–20 mm long with the other two dimensions in the range of 0.01–0.4 mm.

The composition of these crystals was characterized by the energy-dispersive X-ray spectroscopy (EDS) with an Ametek @ EDAX (Model Octane Plus) spectrometer, equipped in a field-emission scanning electron microscope (Hitachi S-4800). Single-crystal X-ray diffraction data were collected from a single crystal coated with mineral oil with Mo K$_\alpha$ radiation, $\lambda = 0.71073$ Å in a Bruker D8 Venture Photon II diffractometer equipped with a multifilm monochromator. Structure refinement was carried out with the program SHELXL-2014/7 implemented in the program suite Apex 3.

The temperature dependence of resistance $R(T)$ was measured with the conventional four-probe method having the current injected along the $a$ axis, the longest dimension of these crystals. The thermopower $S(T)$ measurements were performed by using a homemade setup that was integrated into the commercial Magnetic Property Measurement System (MPMS-3, Quantum Design). The temperature gradient $\Delta T/T \sim 1\%$ was maintained along the $a$ axis during the $S(T)$ measurements, and was recorded with a AuFe/chromel differential thermocouple with an accuracy 0.2%. The precision of thermopower is $\sim 0.5$ µV/K. To investigate the effect of magnetic fields on both $R(T)$ and $S(T)$, we have applied various magnetic fields up to 7 T along the three principal crystallographic axes. The effect of magnetic fields on the thermocouple was found to be negligible.

Our density functional calculations were done using the general potential linearized augmented planewave (LAPW) method[20] as implemented in the WIEN2k code.[21] We used sphere radii of 2.4 bohr for Zr and Hf, and 2.5 bohr for Te, along with local orbitals to treat the semicore states. We used the standard LAPW basis with local orbitals[22] rather than the more efficient APW+lo method,[23] which has larger errors especially away from the linearization energies, potentially of importance for the spin-orbit calculation. We used highly converged basis sets, determined by the criterion $R_{min}k_{max} = 9.0$ ($R_{min} = 2.4$ bohr is the radius of the smallest sphere, $k_{max}$ is the planewave sector cut-off), along with dense zone samplings. The calculations were based on the experimental lattice parameters[2], with all internal atomic coordinates relaxed using the PBE exchange correlation functional.[24] For technical reasons this relaxation was done in a scalar relativistic approximation. We used the resulting crystal structures to calculate the electronic properties and transport coefficients, including spin-orbit. The transport coefficients were obtained using the BoltzTraP code.[25]

## Results and discussions

### 1. Transport properties at zero magnetic field

Fig. 1 displays the temperature dependence of (a) resistance $R(T)$ and (b) thermopower $S(T)$ in the temperature range 2 K < $T$ < 300 K under zero magnetic field for the two different ZrTe$_5$ crystals. The results for S1 and S2 shown here are representative for the single crystals made by different methods, and are found to be reproducible by checking several crystals for each method. For the sample S1 grown with the CVT method, $R(T)$ initially decreases slightly and then exhibits a broad peak centered around $T_p \approx 132$ K, below which a typical metallic behavior was recovered; the thermopower attains a large, positive value ~ 200 µV/K at room temperature, and it undergoes a dramatic sign change crossing zero at $T_p$, (see the dotted line in Fig. 1). All these observations in S1 are consistent with previously published results[3, 5]. Surprisingly, we observed totally different transport behaviors for the sample S2. As seen in Fig. 1, $R(T)$ first decreases upon cooling, showing a metallic behavior, for 200 K< $T$ < 300 K, but then it changes to semiconducting-like behavior by undergoing two successive upturns at about 150 K and 50 K, before reaching a plateau at low temperatures; $S(T)$ remains positive in the whole temperature range without any sign change, yet attains a similar value of ~200 µV/K at room temperature. It is noteworthy that a corresponding slope change of $S(T)$ can be discerned at the characteristic temperatures where the $R(T)$ also exhibits anomalies around 150 K and 50 K.

As mentioned, the resistance peak and the sign reversal in thermopower at $T_p \approx 130$ K in ZrTe$_5$ have been known since 1980s,[3] but a proper understanding of these features remains lacking. The present study implies that these anomalous transport properties are actually sample dependent. Thus, we first clarify the sample differences for these ZrTe$_5$ crystals and then discuss the underlying physics. We performed detailed characterizations on the crystal structure and chemical composition by means of single-crystal X-ray diffraction (SXRD) and energy dispersive spectroscopy (EDS) for the two kinds of crystals.

The SXRD data for both crystals can be well described with the orthorhombic (space group *Cmcm*, No. 63) ZrTe$_5$ structure having Zr and Te1 atoms at 4*c* (0, *y*, ¼) sites, Te2 and Te3 atoms at 8*f* (0, *y*, *z*) positions, respectively. The refinements converged well for both crystals with the average and weighted reliability factors R$_1$= 0.0488(0.0267) and wR$_1$ = 0.1337 (0.0406) for S1(S2), respectively. The smaller R factors of S2 suggest a better crystal quality for those grown with the flux method described above. The detailed information about the structure refinements are given in Table 1. The obtained atomic coordination, occupancies, and selected bond lengths and bond angles are listed in Tables S1 and S2, respectively. As seen in Table 1, the obtained unit-cell parameters, $a$ = 3.9830 (3.9813) Å, $b$ = 14.493 (14.5053) Å, $c$ = 13.700 (13.7030) Å, and $V$ = 790.8 (791.35) Å$^3$ for S1(S2), are very close to those reported previously[1,2], and their differences are less than 0.1%. One of the important implications from the SXRD refinements is that the S1 crystal is slightly Te-deficient with an average composition ZrTe$_{4.86}$, while the S2 crystal has a nearly perfect stoichiometry ZrTe$_{4.98}$. From the obtained site occupancies shown in Table 2, the Te vacancies in S1 are located mainly at the Te2 and Te3 sites, ~3-4%.

The chemical composition and the presence of Te deficiency in S1 are further verified via the EDS measurements on a number of crystals cleaved right before the measurements. The results are summarized in Table S3. The average Zr: Te ratios are 1: 4.60 ± 0.20 and 1: 4.98 ± 0.17 for the S1 and S2, respectively. In general consistent with the SXRD, these results demonstrated unambiguously that a significant amount of Te deficiency is indeed present in the S1 crystals grown with the CVT technique, while the flux method produces ZrTe$_5$ crystals very close to the stoichiometry. These differences should be attributed to the different growth conditions for these two techniques. Although extra tellurium (Te) was added during the CVT growth, the evaporation of Te nonetheless invariably produces sizable Te deficiency; as a matter of fact, Te crystals are observed after the CVT growth[3]. In contrast, crystal growth in Te flux by using the CCS is more effective in achieving the stoichiometric composition.

Based on the above, we can rationalize the distinct transport properties of ZrTe$_5$ shown in Fig. 1 in terms of the different chemical compositions. The transport properties of S2 should be regarded as the intrinsic, or very close to the intrinsic behaviors of ZrTe$_5$; the observed semiconducting-like $R(T)$ is consistent with the first-principles band structure calculations[11] showing a finite gap near the Fermi energy as also detected by the scanning tunneling microscopy measurement[17]. The positive $S(T)$ in the whole temperature range indicated that the Fermi level is slightly below the top of conduction band. On the other hand, the large amount of Te deficiency in the sample S1 introduces electron carriers to the valence band, giving rise to a metallic resistivity and a negative thermopower at low temperatures as observed. As will be shown below, the resistivity peak and the thermopower sign reversal at $T_p$ can be reproduced by our parameter-free transport calculations when taking into account the intrinsic band structure anisotropies at the valence band maximum and conduction band minimum. Our results also provide a simple physical explanation for the contradictory reports in literature regarding the electronic structure of ZrTe$_5$, *i.e.* semiconductor versus Dirac semimetal. [12-18]

2. **Transport properties under magnetic fields**

Despite the apparently different transport properties of these two kinds of ZrTe$_5$ single crystals, a close inspection of the data in Fig. 1 shows that the characteristic anomalies in $R(T)$ and $S(T)$ actually takes place at nearly the same temperature for S1 and S2. For example, the S2 sample shows a broad hump feature in $R(T)$ together with a downturn trend in $S(T)$ near the $T_p$ of S1. This fact implies that these transport property anomalies might have a common origin that is intrinsic to ZrTe$_5$. The similar effects of magnetic fields on the transport properties shown in Figs. 2 and 3 further elaborate this point.

Fig. 2 shows the $R(T)$ and $S(T)$ data of S1 under various magnetic fields up to 7 T applied along the three principal axes. We have plotted the data with the same scale to see clearly the different effect of magnetic field. As shown in Fig. 2(a, b), the effect of magnetic field on $R(T)$ and $S(T)$ is negligible when $H//a$. In contrast, the influence of magnetic field is much enhanced for $H//c$; as seen in Fig. 2(c, d), the most dramatic changes are observed in the temperature range around $T_p$, where the magnitudes for both the resistivity peak and the negative thermopower peak are enhanced by a factor of 2-3. In addition, $T_p$ shifts slightly towards a higher temperature with increasing magnetic field as reported previously[26]. The influence of magnetic field on $R(T)$ and $S(T)$ is the most pronounced for $H//b$, *i.e.* the direction perpendicular to the ZrTe$_3$ sheets (Fig. S1). As seen in Fig. 2(e, f), the resistance peak and the thermopower negative peak become 4-6 times larger under a magnetic field $H = 7$ T. It should be noted that both $R(T)$ and $S(T)$ exhibits obvious enhancement in a wide temperature range well above $T_p$ for this configuration, while the effect is hard to see above $T_p$ for $H//c$.

Similar anisotropic effects of magnetic field on the transport properties of the sample S2 were also observed in Fig. 3. As can be seen, both $R(T)$ and $S(T)$ hardly change under magnetic fields up to 7 T for $H//a$, while they are enhanced significantly when the magnetic field is applied along the $b$ and $c$ axes. For $H//c$, the enhancements appear mainly at temperatures below ~150 K, in particularly, pronounced $S(T)$ peaks reaching as high as ~500 µV/K in concomitant with the wide $R(T)$ plateaus emerge below ~100 K upon the application of magnetic fields. Although similar features are observed for $H//b$, the influence at $T > 200$ K is much stronger than that for $H //c$.

We can thus conclude that the influences of magnetic field on the transport properties of two kinds of ZrTe$_5$ crystals are quite similar, and the strongest effect appears mainly in the vicinity of the characteristic anomalies, regardless the detailed difference of band fillings. Given the 2D character of the crystal and electronic structures, such an anisotropic magnetic field effect on the transport properties can be understood in terms of the electron cyclotron motion in the presence of magnetic fields; the effect is the strongest when $H$ is perpendicular to current applied along the $a$ axis within the ZrTe$_3$ sheets. Moreover, the largest Laudau level interval along the $b$ axis, i.e., $\Delta_{LL}^b \gg \Delta_{LL}^c > \Delta_{LL}^a$, makes the influence of magnetic field appearing in a much more extended temperature range. These results thus suggested that the transport anomalies might originate from an intrinsic, common band structure of ZrTe$_5$, as demonstrated by our first-principles calculations shown below.

3. **Electronic Structure and Transport Calculations**

Density functional calculations were done for ZrTe$_5$ and HfTe$_5$ using the general potential linearized augmented planewave (LAPW) method[20] as implemented in the WIEN2k code[21]. The density of states of both compounds shows hybridized bands derived from Te $p$ and metal $d$ states. This is similar to prior reports.[11] The Te contributions are dominant as may be expected from the stoichiometry and this is also the case near the Fermi levels. The metal $d$ states are more prominent in the conduction bands. The band structure formation is a consequence of Te $p$ states. Spin-orbit is particularly important because these are $p$ states. The complexity of the crystal structure (see Fig. S1) can then be readily understood as due to competition between Te $p$ bonding and accommodation of the metal ions, similar to other complex structure tellurides, such as IrTe$_2$.[27] We note that the separation of the metal ions is ~4 Å, implying that direct interactions will be very weak, and that their coupling will be via the Te lattice.

What is important for transport is the electronic structure near the Fermi energy. We found both compounds to be semiconductors, with indirect band gaps of 0.073 eV for ZrTe$_5$ and 0.047 eV for HfTe$_5$. The density of states of the two compounds near the gap is shown in Fig. 4(a). It is important to note that unlike most topological systems the bands near the valence band maximum (VBM) are very different from those near the conduction band minimum (CBM). In particular, the onset of the density of states is much steeper for electrons at the CBM than for holes at the VBM. The band shapes at the VBM and CBM are also very different. This is seen in Fig. 4(b), which depicts the isosurfaces of the highest valence band 0.05 eV below the VBM and the lowest conduction band 0.05 eV above the VBM. In addition to the pockets near the zone center, the conduction bands show additional pockets in the middle of the zone and at the zone boundaries. It is clearly seen that the shapes of the hole and electron pockets are different, and based on this one may anticipate different transport properties

This is the case. Fig. 5 shows calculated values of the transport function $\sigma/\tau$, where $\sigma$ is conductivity and $\tau$ is the unknown inverse scattering rate; the ratio is determined by the band structure alone and is proportional to the square of the optical Drude plasma frequency, as obtained from the electronic structure for a temperature of 50 K, using the BoltzTraP code [25]. The two compounds behave very similarly from this point of view. The conductivity along the *b*-axis direction (*y*) is very low reflecting the fact that from an electronic transport point of view the material is highly two dimensional. There is also significant anisotropy in the *ac* plane consistent with the report of Tritt and co-workers.[26] Importantly, the high conductivity directions for electrons and holes are opposite, and the hole conductivity is higher than the electron conductivity for transport along the usually measured *a* axis (*x*). Specifically, the hole conductivity is highest along *a*, while the electron conductivity is highest along *c*. This has important implications.

Specifically, $ZrTe_5$ and $HfTe_5$ are usually *n*-type at low temperatures due to Te deficiency as verified in the present work. For a heavily doped narrow gap system, one expects behavior of a degenerate doped (*i.e.* low carrier density metallic) system up to some temperature at which point bipolar conduction starts. At this point the carrier density will increase keeping a balance between electrons and holes, and at high temperature *S(T)* will take the sign of the higher conductivity carriers. One may also anticipate that the resistivity will increase with temperature due to electron phonon scattering below the crossover temperature where bipolar conduction becomes important, and then decrease due to the increase in carrier density. Strong *T* dependence could occur near this crossover due to the exponential factors that appear in the Fermi function. This picture is confirmed by direct calculations of the transport functions.

Fig. 6(a) shows the calculated $S(T)$ along the $a$-axis for different $p$ and $n$-type doping levels. These were obtained in the standard constant scattering time approximation, setting the hole and electron scattering rates equal. As seen, for $n$-type, both $ZrTe_5$ and $HfTe_5$ display a strong doping dependent crossover from electron-like to hole-like $S(T)$ and a relatively constant value at high temperature, as we observed experimentally in Fig. 1. The crossover temperature for $HfTe_5$ is lower than that of $ZrTe_5$, consistent with experimental reports. This difference is due to the smaller band gap of $HfTe_5$. We emphasize that we have not introduced any adjustable parameter to obtain these results. Turning to the value of the $p$-type thermopower at high temperature, it can be seen that our values are somewhat lower than typical experimental values. Such a difference could be corrected by assuming different scattering rates $\tau$ for electrons and holes. While this may be reasonable considering the different nature of the electron and hole bands it would introduce a parameter.

We next turn to the resistivity. We consider the simplest model that follows from the above considerations. Specifically, we take $\sigma/\tau$ as obtained directly from the band structure and adopt a metallic electron-phonon dependence $\tau/1/T$.[28, 29] The inverse of this function is plotted for various doping levels in Fig. 6(b). As can be seen, there are strong peaks in resistivity for low $n$-type doping levels ($\sim 10^{18}$ cm$^{-3}$) at temperatures consistent with experiment and in particular the peak for $HfTe_5$ is at lower temperature than that for $ZrTe_5$. We also note that there is a doping level dependence that explains the sample to sample variation in these materials. The fact that the transport even for $n$-type samples is strongly affected by both holes and electrons at and above the resistivity peak, and that these involve the conductivities in the $a$-axis and $c$-axis directions, but not significantly along $b$, is consistent with the field orientation dependences, specifically the lack of enhancement in the transition only when field is along $a$. Importantly, a semiconducting behavior as seen in Fig. 1(a) for the S2 sample can be reproduced for the undoped material as shown in Fig. 6(c).

Finally, we made a brief comment on thermopower. Normally, thermopower is the most isotropic of the electrical transport quantities, particularly in semiconductors. However, open Fermi surfaces can lead to strongly anisotropic thermopower as in the case of $PdCoO_2$.[30] As seen in Fig. 4(b), the energy isosurfaces become open at low energies. Therefore, it is of interest to examine the transport behavior for the other conduction directions, along *c* (note that there is very little conduction along *b*). We predict that in this direction the behavior of the thermopower is opposite to that along *a*. In specific, there is a change in sign for *p*-type material, and the high temperature saturation for both *n*- and *p*-type is to a negative thermopower. The absolute values are larger for *n*-type than for *p*-type as due to the heavier bands as seen in stronger density of states onset for the conduction bands. The opposite behavior of *S(T)* along *a* and *c* is a consequence of the different high conductivity directions for electrons and holes. It will be of interest to test this prediction should suitable samples become available.

Turning to thermoelectric application, the onset of bipolar conduction is invariably highly detrimental to the figure of merit $ZT \equiv S^2 \sigma T/\kappa$, where *S*, $\sigma$, and $\kappa$ stand for thermopower, electrical conductivity, and thermal conductivity, respectively. It leads to a strong enhancement of $\kappa$ and a decrease of *S*. Thus, for any application it will be important to avoid the bipolar regime. This is defined by the resistivity maximum. Therefore, the best performance would be at cryogenic temperatures below the resistivity maximum, and according to our results for the thermopower would likely be best along the *c*-axis. It is noteworthy that the predicted values of *S* for the naturally occurring *n*-type at 150 K for carrier concentrations of as high as $5 \times 10^{18}$ cm$^{-3}$ exceed 300 $\mu$V/K. Furthermore, the complex band shapes that arise from the spin-orbit splitting[11] are generally favorable for thermoelectric performance.[31] A key issue will be the extent to which the *n*-type doping level can be controlled to achieve appropriate *S* and $\sigma$ in the temperature range below the onset of bipolar conduction.

## Conclusion

In conclusion, we have verified experimentally that ZrTe$_5$ single crystals with Te deficiency show interesting transport properties including the resistance peak and the thermopower sign reversal around $T_p$ = 135 K. In contrast, the nearly stoichiometric ZrTe$_5$ single crystals show *p*-type semiconducting transport behavior at low temperatures. Aided by the first-principles calculations, we have identified a common origin in the peculiar band structures of pentatellurides for these distinct transport behaviors. The longstanding puzzle of the transition in ZrTe$_5$ and HfTe$_5$ can be understood in terms of an onset of bipolar conduction for narrow gap semiconductors with bands having different anisotropies at the VBM and CBM. Given the growing research interesting on pentatellurides from the perspective of topological properties, our present work calls for caution on the sample's stoichiometry in future investigations. Nevertheless, our work also demonstrated that the interesting properties of ZrTe$_5$ can be effectively tailored by controlling carefully the Te content or band filling.

## Acknowledgements

The authors thank Dr. Zhong Wang and Dr. Hongming Weng for enlightening discussions. We are also grateful to Dr. Youting Song for his assistant in the single-crystal X-ray diffraction measurements. J.G.C. is supported by the National Basic Research Program of China (Grant No.2014CB921500), National Science Foundation of China (Grant No. 11574377, 11304371), the Strategic Priority Research Program and the Key Research Program of Frontier Sciences of the Chinese Academy of Sciences (Grant Nos. XDB07020100, QYZDB-SSW-SLH013), and the Opening Project of Wuhan National High Magnetic Field Center (Grant No. 2015KF22) Huazhong University of Science and Technology. D.S.is supported by the Department of Energy, Basic Energy Sciences, through the MAGICS Center, Award DE-SC0014607. J.-Q. Y. is supported by the US Department of Energy, Office of Science, Basic Energy Sciences, Materials Sciences and Engineering Division. D.G.M. acknowledges support from NSF DMR 1410428.

**Table 1. Crystal data and structure refinement for ZrTe$_5$ (S1 and S2) single crystals**

| | S1 (CVT) | S2 (Flux) |
|---|---|---|
| Empirical formula | | |
| Crystal system, space group | Orthorhombic, *Cmcm* | Orthorhombic, *Cmcm* |
| Unit cell dimensions | | |
| $a$ (Å) | 3.9830(7) | 3.9813(3) |
| $b$ (Å) | 14.493(3) | 14.5053(14) |
| $c$ (Å) | 13.700(3) | 13.7030(13) |
| Volume (Å$^3$), Z | 790.8(3), 4 | 791.35(12), 4 |
| Density (calculated) (mg/m$^3$) | 5.977 | 6.121 |
| Formula weight | 711.67(**ZrTe$_{4.86}$**) | 729.22(**ZrTe$_5$**) |
| Temperature (K) | 273(2) | 273(2) |
| Wavelength (Å) | 0.71073 | 0.71073 |
| Absorption coefficient (mm$^{-1}$) | 18.859 | 19.344 |
| F(000) | 1171.0 | 1200.0 |
| Crystal size (mm) | 0.16 ×0.21 × 0.88 | 0.14 ×0.28 × 0.61 |
| θ range for data collection (°) | 2.811-27.606 | 2.809-27.444 |
| Limiting indices | $-5 \leq h \leq 5$, $-18 \leq k \leq 18$, $-17 \leq l \leq 17$ | $-4 \leq h \leq 5$, $-18 \leq k \leq 18$, $-17 \leq l \leq 17$ |
| Reflections collected/unique | 4443 / 538 | 4512 / 544 |
| $R_{int}$ | 0.0586 | 0.0504 |
| Refinement method | Full-matrix least-squares on F$^2$ | Full-matrix least-squares on F$^2$ |
| Data / restraints / parameters | 538 / 0 / 22 | 539 / 0 / 22 |
| Goodness-of-fit on F$^2$ | 1.287 | 1.173 |
| Final R indices [I > 2σ(I)] | $R_1$ = 0.0483, $wR_2$ = 0.1335 | $R_1$ = 0.0215, $wR_2$ = 0.0396 |
| R indices (all data) | $R_1$ = 0.0488, $wR_2$ = 0.1337 | $R_1$ = 0.0267, $wR_2$ = 0.0406 |
| Largest different peak and hole (e Å$^{-3}$) | 3.021, -3.218 | 1.055, -1.303 |

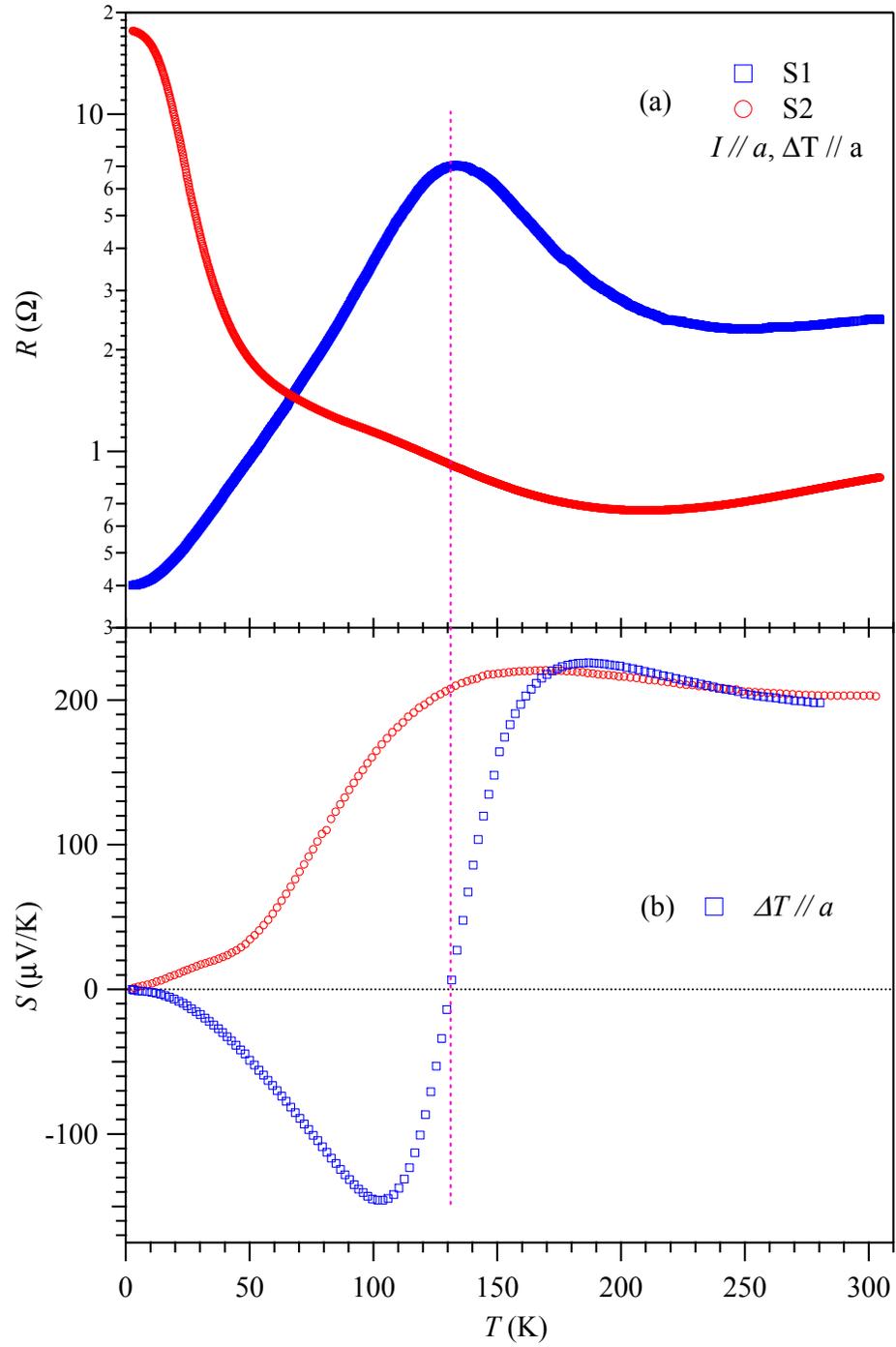

Fig. 1 (Color online) Temperature dependence of (a) resistance $R(T)$ and (b) thermopower $S(T)$ without magnetic field for different $ZrTe_5$ crystals S1 and S2.

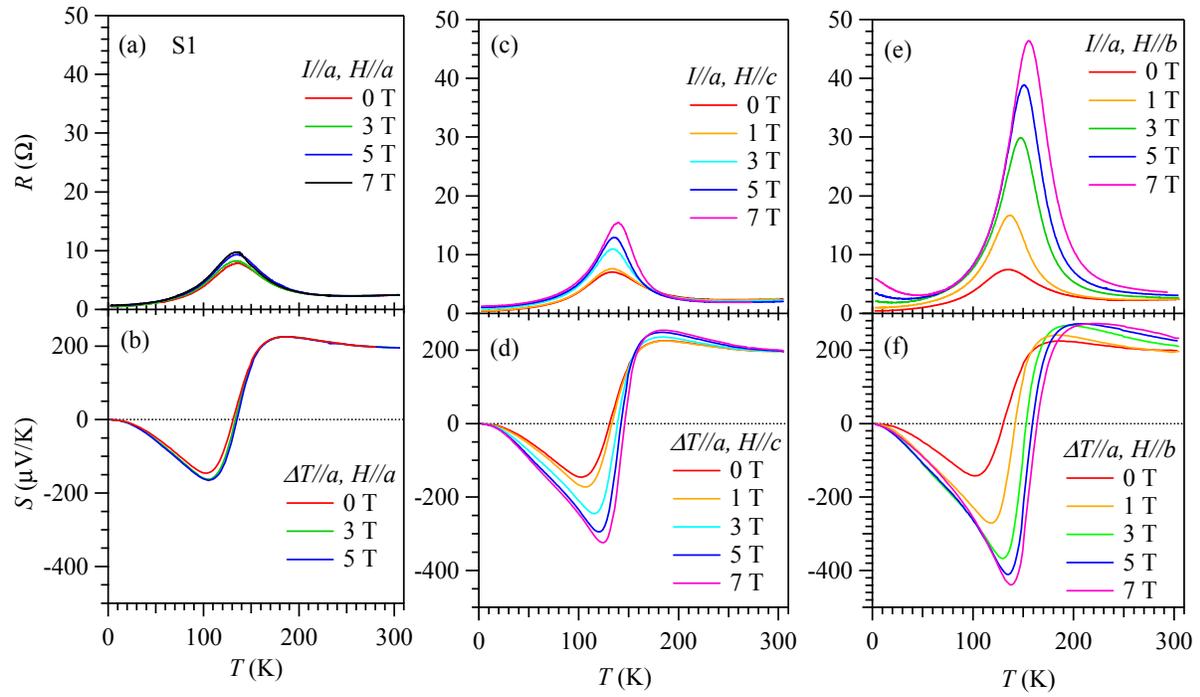

Fig. 2 (Color online) Temperature dependence of resistance $R(T)$ and thermopower $S(T)$ for S1 with the magnetic field applied along the three principal axes.

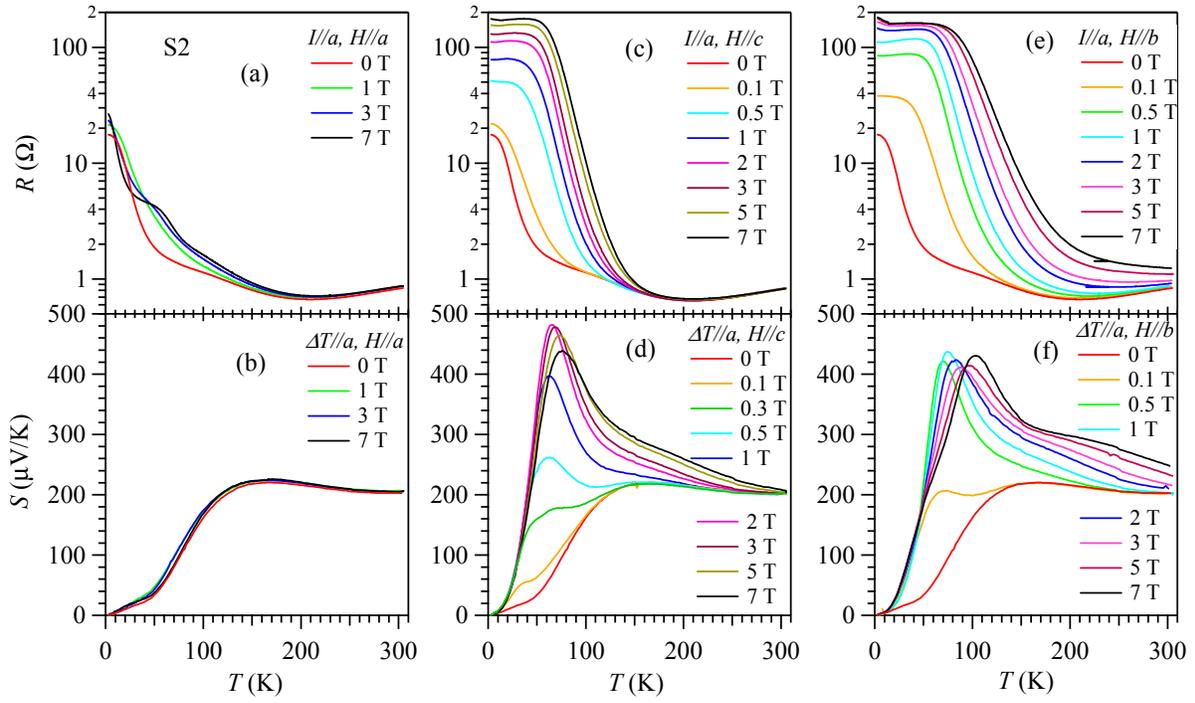

Fig. 3 (Color online) Temperature dependence of resistance $R(T)$ and thermopower $S(T)$ for S2 with the magnetic field applied along the three principal axes.

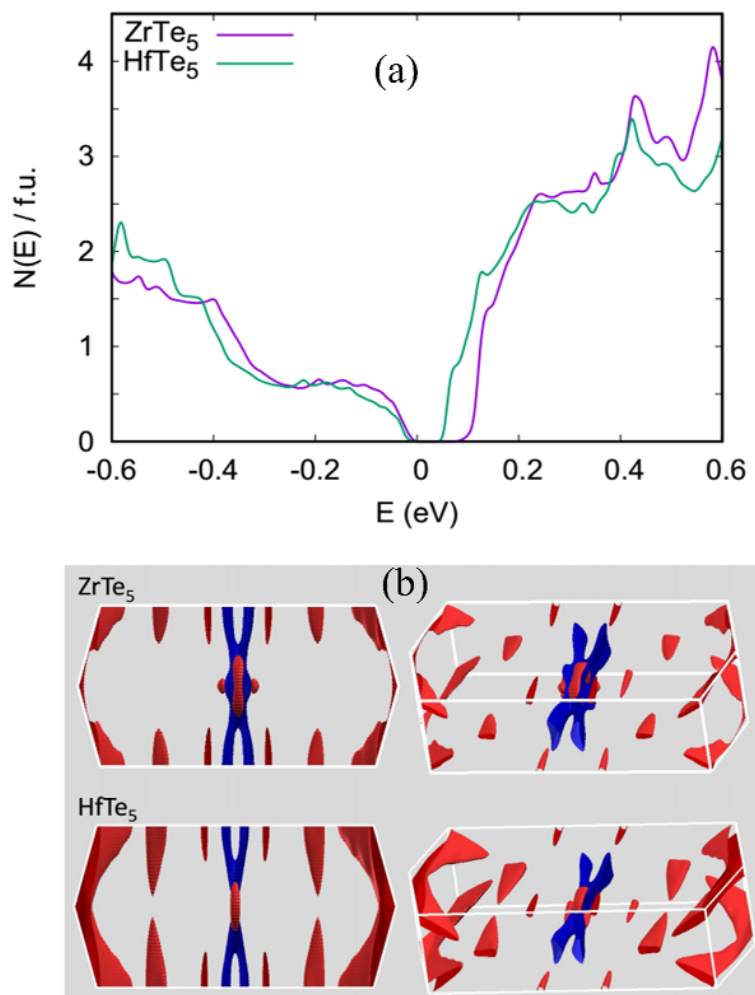

Fig. 4 (Color online) (a): Total electronic DOS of $ZrTe_5$ and $HfTe_5$ on a per formula unit basis. Note the smaller gap in $HfTe_5$ and the asymmetry between electrons and holes. (b) Band energy isosurfaces 0.05 eV from the band edges for $ZrTe_5$ and $HfTe_5$. Hole surfaces below the valence band maximum are in blue, while electron surfaces above the conduction band minimum are in red. Note the very different structure of the hole and electron sheets.

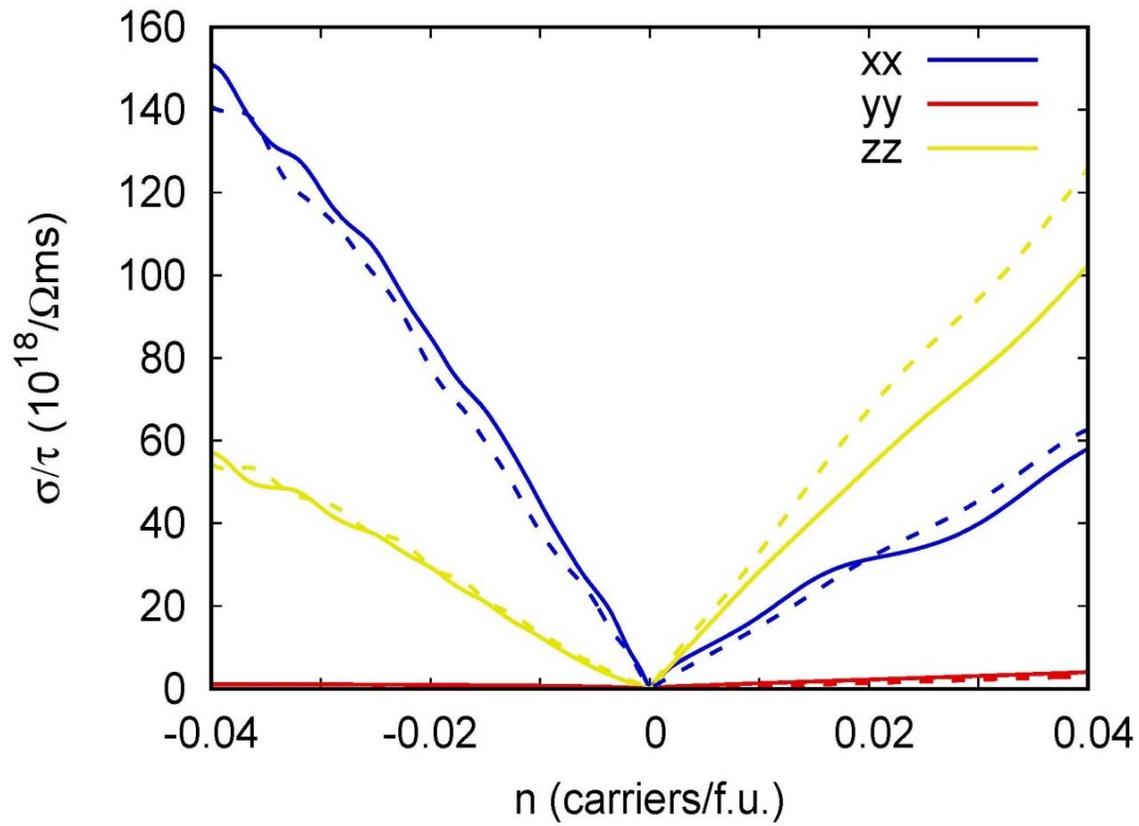

Fig. 5 (Color online) Transport function $\sigma/\tau$ along the three crystallographic directions for $ZrTe_5$ (solid lines) and $HfTe_5$ (dashed lines).

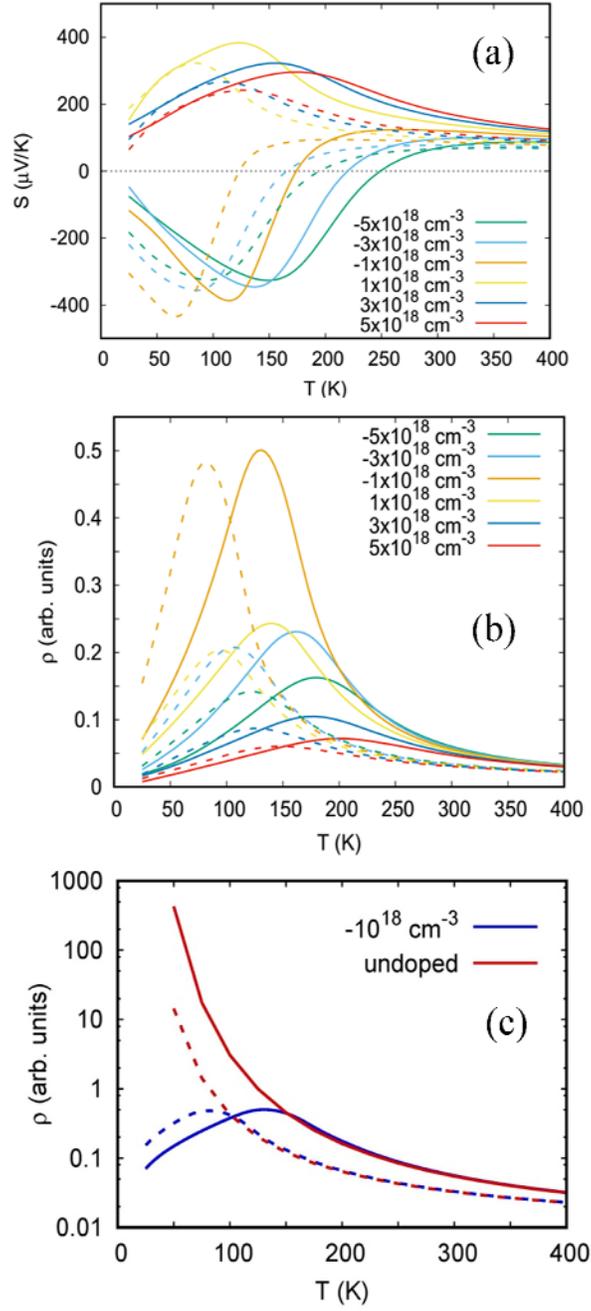

Fig. 6 (Color online) (a) *a*-axis $S(T)$ for various doping levels as obtained in the constant scattering time approximation for $ZrTe_5$ (solid lines) and $HfTe_5$ (dashed lines). (b) *a*-axis resistivity transport function, $\rho$ (see text) for $ZrTe_5$ (solid lines) and $HfTe_5$ (dashed lines) (c) *a*-axis resistivity transport function on a log scale comparing *n*-type doped and undoped material.

Supplementary Materials

# Bipolar Conduction is the Origin of the Electronic Transition in Pentatellurides: Metallic vs. Semiconducting Behavior

P. Shahi[1], D. J. Singh[2*], J. P. Sun[1], L. X. Zhao[1], G. F. Chen[1], H. X. Yang[1], J.-Q. Yan[3], D. G. Mandrus[3,4], H. M. Weng[1], and J.-G. Cheng[1*]

[1.] Beijing National Laboratory for Condensed Matter Physics and Institute of Physics, Chinese Academy of Sciences, Beijing 100190, China
[2.] Department of Physics and Astronomy, University of Missouri, Columbia, MO 65211-7010 USA
[3.] Materials Science and Technology Division, Oak Ridge National Laboratory, Oak Ridge, TN 37831, USA
[4.] Department of Materials Science and Engineering, University of Tennessee, Knoxville, TN 37996, USA

[*]E-mails: singhdj@missouri.edu, jgcheng@iphy.ac.cn

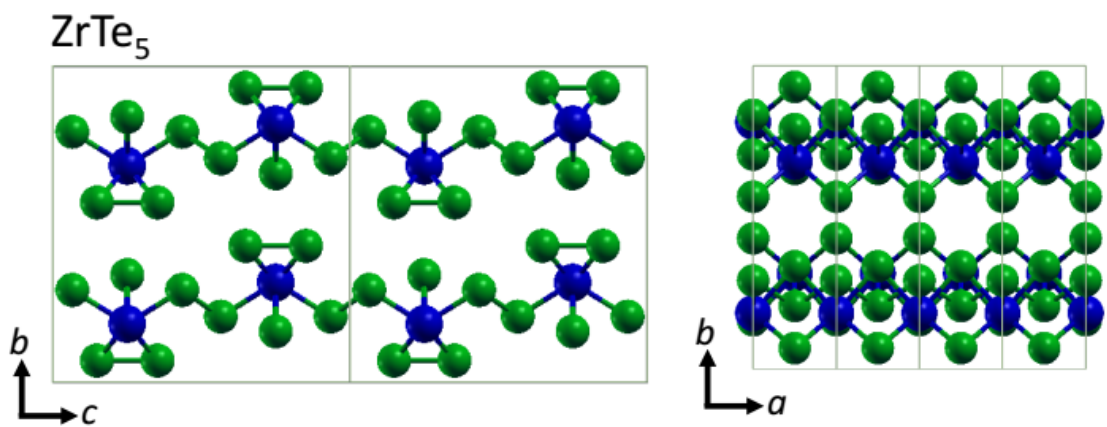

FIG.S1. Structure of $ZrTe_5$ viewed along the *a*-axis (left) and *c*-axis (right) directions. Note the layering along *b*. Zr is shown as blue, while Te is depicted as green.

**Table S1. Atomic coordinates and equivalent isotropic displacement parameters for ZrTe$_5$ (S1 and S2) single crystals. $U_{eq}$ is defined as one third of the trace of the orthogonalized $U_{ij}$ tensor**

| ZrTe$_5$ | Atom | x | y | z | $U_{eq}$ (10$^{-2}$Å$^2$) | Occup. |
|---|---|---|---|---|---|---|
| S1 | Zr | 0.5 | 0.31539(13) | 0.75 | 0.0160(6) | 1 |
| | Te1 | 1 | 0.16350(9) | 0.75 | 0.0174(5) | 0.9907 |
| | Te2 | 0 | 0.43153(6) | 0.64983(7) | 0.0197(5) | 0.9617 |
| | Te3 | 0.5 | 0.20981(7) | 0.56443(7) | 0.0193(5) | 0.9736 |
| S2 | Zr | 0.5 | 0.31538(6) | 0.75 | 0.0125(2) | 1 |
| | Te1 | 1 | 0.16353(4) | 0.75 | 0.01462(17) | 1 |
| | Te2 | 0 | 0.43148(3) | 0.64982(3) | 0.01817(15) | 0.9924 |
| | Te3 | 0.5 | 0.20976(3) | 0.56443(3) | 0.01715(15) | 0.9947 |

| ZrTe$_5$ | | $U_{11}$ | $U_{22}$ | $U_{33}$ | $U_{23}$ | $U_{13}$ | $U_{12}$ |
|---|---|---|---|---|---|---|---|
| S1 | Zr | 0.0137(10) | 0.0193(10) | 0.0151(9) | 0.000 | 0.000 | 0.000 |
| | Te1 | 0.0157(8) | 0.0175(7) | 0.0189(8) | 0.000 | 0.000 | 0.000 |
| | Te2 | 0.0170(7) | 0.0210(6) | 0.0212(6) | 0.0058(4) | 0.000 | 0.000 |
| | Te3 | 0.0188(7) | 0.0240(6) | 0.0151(6) | -0.0014(3) | 0.000 | 0.000 |
| S2 | Zr | 0.0111(5) | 0.0156(5) | 0.0109(4) | 0.000 | 0.000 | 0.000 |
| | Te1 | 0.0126(3) | 0.0154(3) | 0.0159(3) | 0.000 | 0.000 | 0.000 |
| | Te2 | 0.0148(2) | 0.0198(3) | 0.0199(3) | 0.0055(2) | 0.000 | 0.000 |
| | Te3 | 0.0164(2) | 0.0219(3) | 0.0132(2) | -0.0014(2) | 0.000 | 0.000 |

**Table S2. Selected bond lengths and band angles of ZrTe$_5$ (S1 and S2) single crystals**

| | Bond length (Å) | | | Bond angle (°) | |
|---|---|---|---|---|---|
| | S1 | S2 | | S1 | S2 |
| Te1-Zr1 | 2.9685(17) | 2.9689(8) | Zr1-Te1-Zr1$^c$ | 84.27(6) | 84.21(3) |
| Te1-Zr1$^c$ | 2.9685(17) | 2.9488(8) | Te2$^d$-Te2-Zr1$^b$ | 62.24(2) | 62.234(10) |
| Te2-Te2$^d$ | 2.745(2) | 2.7456(9) | Te2$^d$-Te2-Zr1 | 62.24(2) | 62.234(10) |
| Te2-Zr1$^b$ | 2.9466(13) | 2.9468(6) | Zr1$^b$-Te2-Zr1 | 85.04(5) | 84.99(2) |
| Te2-Zr1 | 2.9466(13) | 2.9468(6) | Te3$^f$-Te3-Te3$^g$ | 86.55(5) | 86.48(2) |
| Te3-Te3$^f$ | 2.9050(14) | 2.9058(6) | Te3$^f$-Te3-Zr1 | 108.29(5) | 108.25(2) |
| Te3-Te3$^g$ | 2.9050(14) | 2.9058(6) | Te3$^g$-Te3-Zr1 | 108.29(5) | 108.25(2) |
| Te-Zr1 | 2.9672(14) | 2.9686(7) | Te2-Zr1-Te2$^e$ | 110.33(7) | 110.29(3) |
| Zr1-Te2$^e$ | 2.9466(13) | 2.9468(6) | Te2-Zr1-Te2$^c$ | 85.04(5) | 84.99(2) |
| Zr1-Te2$^c$ | 2.9466(13) | 2.9468(6) | Te2$^e$-Zr1-Te2$^c$ | 55.52(4) | 84.99(2) |
| Zr1-Te2$^d$ | 2.9466(13) | 2.9468(6) | Te2-Zr1-Te2$^d$ | 55.52(4) | 55.53(2) |
| Zr1-Te3$^e$ | 2.9672(14) | 2.9686(7) | Te2$^e$-Zr1-Te2$^d$ | 85.04(5) | 84.99(2) |
| Zr1-Te1$^b$ | 2.9684(17) | 2.9689(8) | Te2$^c$-Zr1-Te2$^d$ | 110.33(7) | 110.29(3) |
| | | | Te2-Zr1-Te3$^e$ | 133.915(17) | 133.946(8) |
| | | | Te2$^g$-Zr1-Te3$^e$ | 84.00(3) | 84.026(12) |
| | | | Te2$^c$-Zr1-Te3$^e$ | 133.916(17) | 133.945(8) |
| | | | Te2$^d$-Zr1-Te3$^e$ | 84.00(3) | 84.026(12) |
| | | | Te2-Zr1-Te3 | 84.00(3) | 84.024(12) |
| | | | Te2$^e$-Zr1-Te3 | 133.916(18) | 133.945(8) |
| | | | Te2$^c$-Zr1-Te3 | 84.00(3) | 84.025(12) |
| | | | Te2$^d$-Zr1-Te3 | 133.916(18) | 133.945(8) |
| | | | Te3$^e$-Zr1-Te3 | 117.92(7) | 117.86(3) |
| | | | Te2-Zr1-Te1$^b$ | 88.29(2) | 88.342(12) |
| | | | Te2$^e$-Zr1-Te1$^b$ | 151.29(3) | 151.278(15) |
| | | | Te2$^c$-Zr1-Te1$^b$ | 151.29(3) | 151.278(15) |
| | | | Te2$^d$-Zr1-Te1$^b$ | 88.29(2) | 88.343(12) |
| | | | Te3$^e$-Zr1-Te1$^b$ | 67.52(3) | 67.488(16) |
| | | | Te3-Zr1-Te1$^b$ | 67.52(3) | 67.489(16) |
| | | | Te2-Zr1-Te1 | 151.29(3) | 151.277(15) |
| | | | Te2$^e$-Zr1-Te1 | 88.29(2) | 88.343(12) |
| | | | Te2$^c$-Zr1-Te1 | 88.29(2) | 88.343(12) |
| | | | Te2$^d$-Zr1-Te1 | 151.29(3) | 151.278(15) |
| | | | Te3$^e$-Zr1-Te1 | 67.52(3) | 67.489(16) |
| | | | Te3-Zr1-Te1 | 67.52(3) | 67.489(16) |
| | | | Te1$^b$-Zr1-Te1 | 84.27(6) | 84.21(3) |

Symmetry transformations used to generate equivalent atoms: (b) -1+x, +y, +z; (c) 1+x, +y, +z; (d) –x, +y, 3/2-z; (e) 1-x, +y, 3/2-z; (f) ½-x, ½-y, 1-z; (g) 3/2-x, ½-y, 1-z.

**Table S3. A summary of the EDS results on ZrTe$_5$ (S1 and S2) crystals**

| ZrTe$_5$ | Sample No. | Zr | Te |
|---|---|---|---|
| S1 | #1 | 18.37 | 81.63 |
| | #2 | 18.93 | 81.07 |
| | #3 | 17.98 | 82.02 |
| | #4 | 16.61 | 83.39 |
| | #5 | 17.82 | 82.18 |
| | #6 | 18.26 | 81.74 |
| | #7 | 18.48 | 81.52 |
| | #8 | 16.47 | 83.26 |
| | Average | 17.90 | 82.10 |
| | Zr:Te | 1 | 4.60± 0.20 |
| S2 | #1 | 16.46 | 83.54 |
| | #2 | 16.14 | 83.86 |
| | #3 | 17.0 | 83.0 |
| | #4 | 18.03 | 91.97 |
| | #5 | 17.75 | 82.25 |
| | Average | 17.08 | 84.92 |
| | Zr:Te | 1 | 4.98± 0.17 |